\documentclass[journal,10pt,twocolumn]{IEEEtran}

\usepackage{amsmath}
\usepackage{hyperref}
\usepackage[utf8]{inputenc}
\usepackage{listings}
\usepackage{multirow}
\usepackage{graphicx}
\usepackage{soul}
\usepackage{subfigure}
\usepackage[sort,square,numbers]{natbib}

\usepackage{tikz}
\usepackage{tikz-qtree}
\usetikzlibrary{trees}
\usepackage[capitalise]{cleveref}

\usepackage[inline,nomargin,marginclue,status=draft]{fixme}
\fxsetup{envlayout=color,theme=color}
\FXRegisterAuthor{db}{adb}{DB}
\FXRegisterAuthor{ds}{ads}{DS}
\FXRegisterAuthor{mi}{ami}{MI}
\FXRegisterAuthor{ap}{aap}{AP}

\newcommand{\Samsung}{\textit{Samsung}}
\newcommand{\Apple}{\textit{Apple}}

\newcommand{\Huawei}{\textit{Huawei}}

\newcommand{\LG}{\textit{LG}}

\newcommand{\Exiftool}{\textit{Exiftool}}
\newcommand{\Kdenlive}{\textit{Kdenlive}}
\newcommand{\Avidemux}{\textit{Avidemux}}
\newcommand{\APremiere}{\textit{Adobe Premiere}}
\newcommand{\Premiere}{\textit{Premiere}}
\newcommand{\Ffmpeg}{\textit{ffmpeg}}
\newcommand{\sk}{\textit{EVA-7K}}
\newcommand{\LT}{\textit{EVA}}

\tikzset{
    containernode/.style = {shape=rectangle, rounded corners, draw,
        top color=white, bottom color=gray!40, anchor=west},
    containerfield/.style = {containernode, top color=white, bottom color=gray!5}
}

\title{Efficient video integrity analysis through container characterization}

\author{Pengpeng~Yang,~\IEEEmembership{Student~Member,~IEEE,}
        Daniele~Baracchi,
        Massimo~Iuliani,
        Dasara~Shullani,
        Rongrong~Ni,
        Yao~Zhao,~\IEEEmembership{Senior~Member,~IEEE,}
        and~Alessandro~Piva,~\IEEEmembership{Fellow,~IEEE}%
\thanks{This work was supported in part by the National Key Research and Development of China (No. 2016YFB0800404), the National NSF of China (Nos. 61672090, 61532005, U1936212) and the Fundamental Research Funds for the Central Universities (Nos. 2018JBZ001, 2017YJS054). It was also supported in part by the Air Force Research Laboratory and in part by the Defense Advanced Research Projects Agency under Grant FA8750-16-2-0188. Finally it was supported by the Italian
 Ministry of Education, Universities and Research MIUR under Grant 2017Z595XS. }%
\thanks{P. Yang, R. Ni, and Y. Zhao are with the Beijing Key Laboratory of Advanced Information Science and Network Technology and the Institute of Information Science, Beijing Jiaotong University, Beijing 100044, China.}%
\thanks{D. Baracchi, M. Iuliani, D. Shullani, and A. Piva are with the Department of Information Engineering, University of Florence, Via di S. Marta, 3, 50139 Florence, Italy.}%
\thanks{M. Iuliani, and A. Piva are with the FORLAB, Multimedia Forensics Laboratory, PIN Scrl, Piazza G. Ciardi, 25, 59100 Prato, Italy.}%
\thanks{Please address correspondence to Rongrong Ni (email: rrni@bjtu.edu.cn) and Alessandro Piva (email: alessandro.piva@unifi.it).}%
\thanks{Digital Object Identifier 10.1109/JSTSP.2020.3008088}%
\thanks{© 2020 IEEE. Personal use of this material is permitted. Permission from IEEE must be obtained for all other uses, in any current or future media, including reprinting/republishing this material for advertising or promotional purposes, creating new collective works, for resale or redistribution to servers or lists, or reuse of any copyrighted component of this work in other works.}}

\begin{document}

\maketitle

\begin{abstract}
    Most video forensic techniques look for traces within the data stream that are, however, mostly ineffective when dealing with strongly compressed or low resolution videos.
    Recent research highlighted that useful forensic traces are also left in the video container structure, thus offering the opportunity to understand the life-cycle of a video file without looking at the media stream itself.

    In this paper we introduce a container-based method to identify the software used to perform a video manipulation and, in most cases, the operating system of the source device.
    As opposed to the state of the art, the proposed method is both efficient and effective and can also provide a simple explanation for its decisions.
    This is achieved by using a decision-tree-based classifier applied to a vectorial representation of the video container structure.
    We conducted an extensive validation on a dataset of $7000$ video files including both software manipulated contents (\Ffmpeg{}, \Exiftool{}, \APremiere{}, \Avidemux{}, and \Kdenlive{}), and videos exchanged through social media platforms (Facebook, TikTok, Weibo and YouTube). 
    This dataset has been made available to the research community.
    The proposed method achieves an accuracy of $97.6 \%$ in distinguishing pristine from tampered videos and classifying the editing software, even when the video is cut without re-encoding or when it is downscaled to the size of a thumbnail.
    Furthermore, it is capable of correctly identifying the operating system of the source device for most of the tampered videos.
\end{abstract}


\section{Introduction}
\label{sec:intro}
Digital videos are becoming more and more relevant in the communication among users and in providing information.
Recent statistics show that the current global average of video consumption per day stands at 84 minutes and it is expected to increase and hit 100 minutes per day by 2021\footnote{https://www.oberlo.com/blog/video-marketing-statistics accessed on March 2020.}.
Therefore, it is not surprising that digital videos are often involved in investigations and other forensic analysis.
At the same time, video editing programs, both open source (e.g. \Ffmpeg{}) and commercial (e.g. \APremiere{}), allow users to easily cut and manipulate videos to create fake contents.

Video Forensics develops algorithms for assessing video integrity and authenticity by looking at the digital traces left during the video life-cycle~\cite{milani2012overview}.
Most of the existing video forensic techniques verify the authenticity of a video file by investigating the presence of inconsistencies in pixel statistics.
For example, double encoding or manipulation can be detected by analyzing prediction residuals~\cite{Shanableh13} or macroblock types~\cite{vazquez2012,gironi2014,vazquez2019video}.
Similarly, traces of frame rate up-conversion can be used to prove malicious video processing~\cite{ding2017identification,xia2017detecting}.
Recent works have also successfully employed deep-learning techniques to detect video forgeries~\cite{d2017autoencoder}.
Jamimamul et al.~\cite{bakas2018digital} focused on inter–frame video forgery detection by designing a 3D Convolutional neural network.
Verde et al.~\cite{verde2018video} introduced a CNN-based approach to detect and localize splicing manipulation by learning video codec traces.

A major drawback of most of those techniques is their high computational cost; furthermore, strong compressions and downsampling often hide forensic traces, thus severely restricting the number of scenarios where those methods can be employed.

Recently, a new research branch highlighted that video integrity\footnote{Note that integrity and authenticity are different concepts.
Integrity is proved when the imagery is complete and unaltered, from the time of acquisition or generation through the life of the imagery; indeed, content authentication is used to determine whether the visual content depicted in imagery is a true and accurate representation of subjects and events.
More details can be found on the Best Practices for Image Authentication of the Scientific Working Group on Digital Evidences~\cite{SWGDE}.} can be determined using information hidden in the whole video file and not just in the video stream~\cite{Gloe2014S68}.
Video files, in fact, are written to disk using a specific structure called container, comprising multiple streams (video, audio, subtitles) and metadata, which are exploited by decoding software to correctly reproduce the video.
Guera et al.~\cite{gueera2019we} showed how to identify forged videos without looking at the pixel space.
To do so, they extracted high level features (multimedia stream descriptors) related to video coding using \emph{ffprobe}. 
However, the tree-shaped container structure is not taken into account, thus discarding most of the overall available information.
Iuliani et al.~\cite{iuliani2018video} highlighted that video integrity can be assessed by looking at the video file container structure, since any post-processing operation alter the content and the position of some atoms and field-values.
This approach turned out to be also promising in classifying the source brand of native videos.

However, \citet{iuliani2018video} merely detects a loss of integrity, without providing a human-interpretable explanation of the reasoning behind its decisions.
Furthermore, the method has a linear computational cost since it requires to check the dissimilarity of the probe video with all available reference containers.
As a consequence, an increase of the reference dataset size leads to a higher computational effort.
Furthermore, both \cite{gueera2019we, iuliani2018video} do not provide any characterization of manipulated videos, nor any explainability of the achieved outcome.

In this paper we introduce an efficient method for the analysis of video file containers that allows both to characterize identified manipulations and to provide an explanation for the outcome.
The proposed approach is based on Decision Trees~\cite{quinlan1986induction}, a non-parametric learning method used for classification problems in many signal processing fields.
Their key feature is the ability to break down a complex decision-making process into a collection of simpler decisions.
We enriched the tool with a likelihood ratio framework designed to automatically clean up the container elements that only contribute to source intra-variability.

With respect to the state of the art, the proposed method, simply called \LT{} from now on, offers new forensic opportunities, such as: identifying the manipulating software (e.g. \APremiere{}, \Ffmpeg{}, \dots); providing additional information related to the original content history, such as the source device operating system.

The process is extremely efficient since a decision can be taken by checking the presence of a small number of features, independently on the video length or size.
Furthermore, \LT{} can provide a simple explanation for the process leading to an outcome, since container symbols used to take a decision can be inspected.
To the best of our knowledge, this is the first video forensic method with all these desirable traits.
Experiments have been performed using videos produced by $34$ modern smartphones from some of the most popular brands on the market, e.g. \Apple{}, \Samsung{}, \LG{}, \Huawei{}.
Tampered contents were generated using both automated processing and manual user operations.
This approach allowed us to build a sizeable, realistic dataset.
Manipulations include contents generated using \Exiftool{}, \Ffmpeg{}, \APremiere{}, \Avidemux{} and \Kdenlive{}.

Eventually, we investigated whether a container-based approach is effective when dealing with videos exchanged through Facebook, TikTok, Weibo, and YouTube.
Overall, the experimental validation involved seven thousands videos.
This paper is organised as follows: \cref{sec:file_format} describes the video container standard; \cref{sec:theory} introduces the mathematical tools to represent and analyse the video file container; \cref{sec:experiments,sec:comparison} are devoted to the experimental validation of the proposed techniques; finally, \cref{sec:conclusions} draws some final remarks and outlines future works.

\begin{figure}[t]
    \begin{tikzpicture}[%
        grow via three points={%
            one child at (0.5,-0.6) and
            two children at (0.5,-0.6) and (0.5,-1.2)
        },
        every node/.style = {font=\small},
        atom/.style = {containernode},
        field/.style = {containerfield},
        ellipsis/.style = {},
        edge from parent path={
            (\tikzparentnode.south) |- (\tikzchildnode.west)
        }]
        \node [atom] {root}
        child { node [atom] {ftyp}
            child { node [field] {majorBrand: isom} }
            child { node [field] {minorVersion: 0} }
            child { node [field] {compatibleBrand\_1: isom} }
            child { node [field] {compatibleBrand\_2: 3gp4} }
            child { node [ellipsis] {...} }
        }
        child [missing] {}
        child [missing] {}
        child [missing] {}
        child [missing] {}
        child [missing] {}
        child { node [atom] {moov}
            child { node [field] {stuff: MovieBox} }
            child {
                node [atom] {mvhd}
                child {
                    node [field] {creationTime: Thu Aug 06...}
                }
                child {
                    node [field] {modificationTime: Thu Aug 06...}
                }
                child {
                    node [field] {timescale: 1000}
                }
                child {
                    node [field] {duration: 73432}
                }
                child {
                    node [ellipsis] {...}
                }
            }
            child [missing] {}
            child [missing] {}
            child [missing] {}
            child [missing] {}
            child [missing] {}
            child {
                node [atom] {trak}
                child { node [ellipsis] {...} }
            }
            child [missing] {}
            child {
                node [atom] {trak}
                child { node [ellipsis] {...} }
            }
            child [missing] {}
            child {
                node [ellipsis] {...}
            }
        }
        child [missing] {}
        child [missing] {}
        child [missing] {}
        child [missing] {}
        child [missing] {}
        child [missing] {}
        child [missing] {}
        child [missing] {}
        child [missing] {}
        child [missing] {}
        child [missing] {}
        child [missing] {}
        child { node [atom] {mdat}
            child { node [ellipsis] {...} }
        }
        ;
    \end{tikzpicture}
    \caption{Pictorial representation of an MP4-like video container structure.}\label{fig:samplecontainer}
\end{figure}
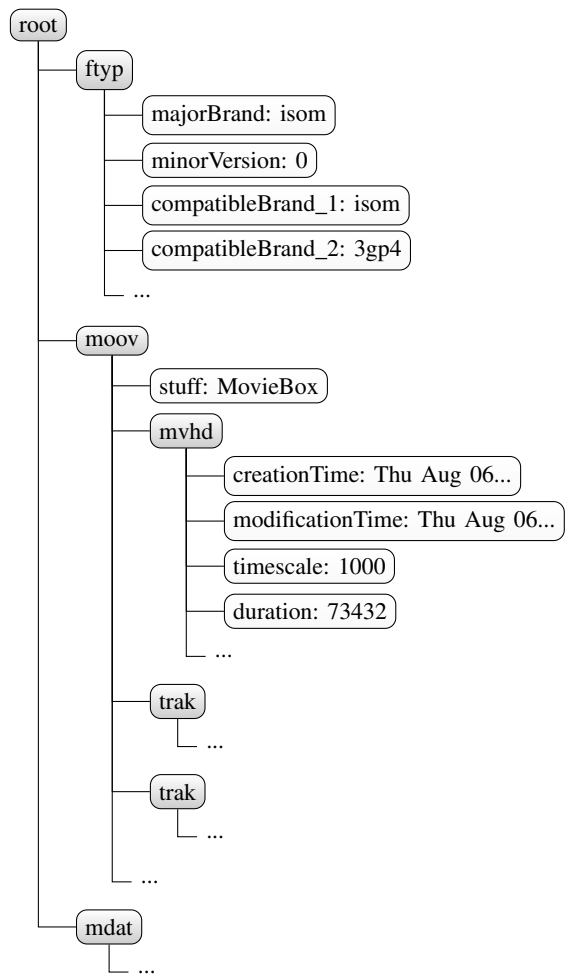

\section{Video File Format}
\label{sec:file_format}
Most smartphones and compact cameras output videos in \texttt{mp4}, \texttt{mov}, or \texttt{3gp} format.
These video packaging refer to the same standard, ISO/IEC 14496 Part 12~\cite{iso}, that defines the main features of \textit{MP4}~\cite{mp4} and \textit{MOV}~\cite{mov} containers while leaving a wide margin for those who implement it.
In~\cref{fig:samplecontainer} we provide an example of an MP4-like container, a tree-like structure describing the video file with respect to three aspects: how the bytes are organized (physical aspect); how the audio/video streams are synchronized (temporal aspect); and how the latter two aspects are linked (logical aspect).

Each node (\emph{atom}) is identified by a unique 4-byte code.
It consists of a header which describes its role in the container and possibly some associated data.
The first atom to appear in a container has to be \texttt{ftyp}\footnote{The reader can refer to \url{http://www.ftyps.com/} for further details.}, since it defines the best usage and compatibility of the video content.
The video structural information is separated from the data itself, indeed the first one is stored in the movie atom (\texttt{moov}) and the second one in the media data atom (\texttt{mdat}).
The \texttt{moov} atom links the logical and timing relationships of the video-samples, and provides pointers to their \texttt{mdat} location.
It is worth noting that the \texttt{moov} sub-tree can contain one or more \texttt{trak} atoms, depending on the number of streams present in a video (i.e. visual-stream and/or audio-stream).

\section{Proposed Approach}
\label{sec:theory}
We can represent a video container as a labelled tree where internal nodes and leaves correspond to, respectively, atoms and field-value attributes.
A video container $X$ can be characterised by the set of symbols $\{ s_1, \dots s_m \}$, where $s_i$ can be: (i) the path from the root to any field (value excluded), also called \textit{field-symbols}; (ii) the path from the root to any field-value (value included), also called \textit{value-symbols}.
An example of this representation can be\footnote{Note that $@$ is used to identify atom parameters and \texttt{root} is used for visualization purpose but it is not part of the container data.}:
\begin{description}
\item[$s_1$]= $[$\texttt{ftyp/$@$majorBrand}$]$
\item[$s_2$]= $[$\texttt{ftyp/$@$majorBrand/isom}$]$
\item[\dots]
\item[$s_i$]= $[$\texttt{moov/mvhd/$@$duration}$]$
\item[$s_{i+1}$]= $[$\texttt{moov/mvhd/$@$duration/73432}$]$
\item[\dots]
\end{description}

Overall, we denote with $\Omega$ the set of all unique symbols $s_1, \dots, s_M$ available in the world set of digital video containers $\mathcal{X}=\{X_1, \dots, X_N \}$.
Similarly, $\mathcal{C} = \{C_1, \dots, C_s\}$ denotes a set of possible origins (e.g., \Huawei{} P9, \Apple{} iPhone 6s).
Given a container $X$, the different structure of its symbols $\{ s_1,\dots, s_m \}$ can be exploited to assign the video to a specific class $C_u$.

\begin{figure}
    \centering
    \begin{tikzpicture}
        \tikzset{level distance=60pt}
        \tikzset{sibling distance=30pt}
        \tikzset{edge from parent/.style={draw,edge from parent path={(\tikzparentnode.south)-- +(0,-8pt)-| (\tikzchildnode)}}}
        \Tree [.\node[draw, rounded corners, fill=gray!10]{count(root/moov/udta/XMP\_/@stuff) $\leq$ 0.5};
                [.\node[draw, rounded corners, fill=gray!5]{class=Native-iOS};] [.\node[draw, rounded corners, fill=gray!5]{class=Exiftool-iOS};] ]
    \end{tikzpicture}
    \caption{%
        Decision tree applied to distinguish iOS native videos from iOS videos with modified dates through \Exiftool{}.
        The decision is easily explainable since it is taken by simply looking at the presence of the symbol \texttt{moov/udta/XMP\_/@stuff}.
    }\label{fig:tree_exiftool}
\end{figure}
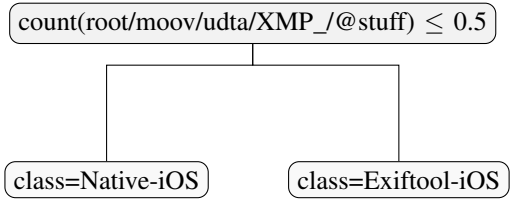

For this purpose binary decision trees~\cite{safavian1991survey} are employed to build a set of hierarchical decisions.
In each internal tree node the input data is tested against a specific condition; the test outcome is used to select a child as the next step in the decision process.
Leaf nodes represent decisions taken by the algorithm.
An example is reported in Fig.~\ref{fig:tree_exiftool}.
More specifically, in our approach we adopted the growing-pruning-based Classification And Regression Trees (CART)~\cite{breiman2017classification}.

Given the size of unique symbols $|\Omega| = M$, a video container $X$ is converted into a vector of integers $X \mapsto (x_1 \dots x_M)$ where $x_i$ is the number of times that $s_i$ occurs into $X$.
This approach is inspired by the bag-of-words representation~\cite{schutze2008introduction} used to reduce variable-length documents to a fixed-length vectorial representation.

\begin{figure}
   {\includegraphics[width=\linewidth]{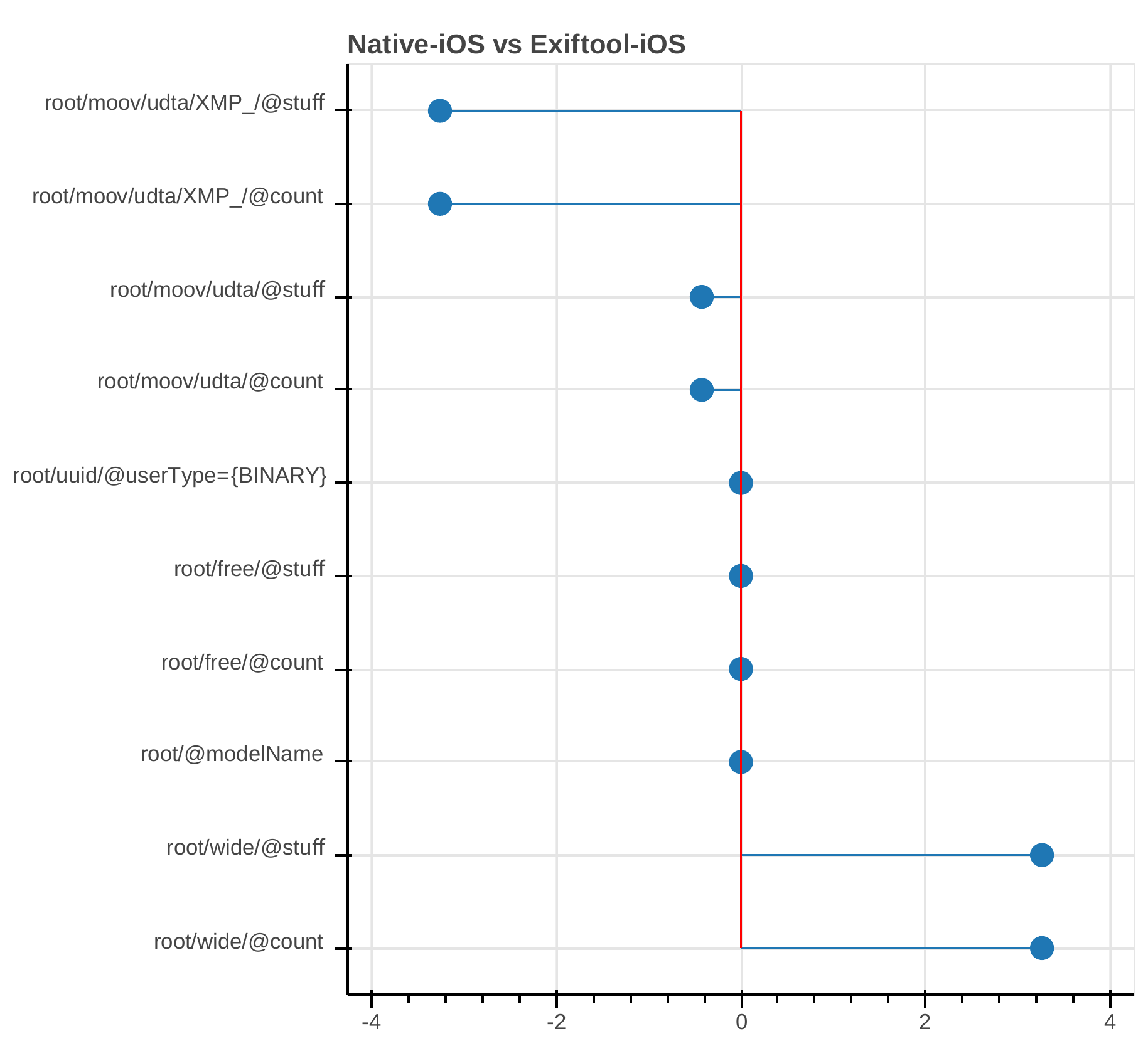}}
\caption{ LLRs of symbols obtained when comparing native videos with ones altered through \Exiftool{}. Values far from $0$ are automatically included in the analysis. Values close to $0$ in all compared classes are excluded from the analysis. Note that @ is used to identify a tree leaf.} \label{fig:LLR_exiftool}
\end{figure}

Note that $X$ contains several symbols that are not representative of any class, thus contributing to class intra-variability only (e.g. information related to video length, acquisition date and time).
This information is expected to introduce noise in the decision process and it should be possibly removed.
Thus, we pre-filtered the data in $\Omega$ by using the likelihood ratio framework.
Given two classes $C_u, C_v, u \neq v$, and a symbol $s_i$, the log-likelihood ratio (LLR)
\begin{equation}
    \label{eq:LR}
    \log L_{u,v}(s_i) = \log \frac{P(s_i | C_u)}{P(s_i | C_v)}
\end{equation}
is computed by approximating the conditional probabilities
\begin{align*}
    P(s_i | C_u) = W_{C_u}(s_i)\\
    P(s_i | C_v) = W_{C_v}(s_i),
\end{align*}
where $W_{C_u}(s_i)$ and $W_{C_v}(s_i)$ are the frequencies of $s_i$ in ${C_u}$ and $C_v$ respectively\footnote{We avoid null frequencies by adding one to both the numerator and the denominator.}.
The symbol $s_i$ is preserved only if $\exists \ u,v, \ u \neq v : \log L_{u,v}(s_i) > \tau$, with $\tau$ a threshold, otherwise it is considered useless and then removed from $\Omega$.
It should be noted that using the likelihood ratio we can possibly keep a \textit{field-symbol} while discarding its corresponding \textit{value-symbol}, or vice-versa.
In this way we can automatically understand whether the value or the field are relevant for the classification.
As an example, we consider two classes:
\begin{itemize}
\item[] $C_u$: iOS devices, native videos;
\item[] $C_v$: iOS devices, dates modified through \Exiftool{} (see Section~\ref{sec:experiments} for details).
\end{itemize}
In~\cref{fig:LLR_exiftool} some achieved LLRs are reported.
The symbols \texttt{moov/udta/XMP\_/@stuff}, \texttt{moov/udta/XMP\_/@count}, \texttt{wide/@stuff}, \texttt{wide/@count} are clearly relevant in identifying this kind of operation on devices equipped with iOS.
On the other hand, symbols like \texttt{free/@stuff} will be possibly filtered since their LLR is close to zero.
In this case, the manipulation only affects a small set of symbols.
Indeed, the decision tree can detect such a processing in a single step, by looking, for instance, at the presence of \texttt{moov/udta/XMP\_/@stuff}, as shown in~\cref{fig:tree_exiftool}.

\section{Integrity Verification}
\label{sec:comparison}
The first relevant experimental question is whether the proposed approach is capable of distinguishing between pristine and tampered videos.
To answer that we created a new collection of videos, starting from VISION~\cite{vision}, a publicly available dataset that includes native videos from $35$ smartphones of $10$ different brands.
As it would have not been feasible to perform the editing operations, upload, and download of all the videos in VISION, we selected $4$ videos for each device, thus obtaining a total of $140$ pristine videos.
Then, we created $1260$ {($140 \times 9$ editing operations) tampered videos, both automatically generated with \Ffmpeg{} and \Exiftool{}, and manually created through \Kdenlive{}, \Avidemux{} and \APremiere{}.
More specifically:
\begin{itemize}
    \item \textbf{cut with re-encoding:} each video was cut through \Ffmpeg{} and re-encoded\footnote{The operation is performed with \Ffmpeg{} version 3.4.6 through the command \texttt{ffmpeg -i \$file -ss 00:00:03 -t 00:00:05 -vcodec libx264 -acodec copy \$name}};
    \item \textbf{cut without re-encoding:} each video was cut through \Ffmpeg{} by copying the audio and video coding parameters\footnote{The operation is performed with the command \texttt{ffmpeg -i \$file -ss 00:00:03 -t 00:00:05 -c copy \$name}} to minimize the traces left by the operation;
    \item \textbf{speed up:} each video was speeded up\footnote{The operation is performed through the command \texttt{ffmpeg -i \$file -vf "setpts=0.25*PTS" \$name} for all the other devices.} through \Ffmpeg{};
    \item \textbf{slow down:} each video was cut through \Ffmpeg{} and slowed down\footnote{The operation is performed with the command \texttt{ffmpeg -i \$file -ss 00:00:03 -t 00:00:15 -vf "setpts=4*PTS" \$name}};
    \item \textbf{cut + downscale:} each video was cut through \Ffmpeg{} and downscaled\footnote{The operation is performed on with the command \texttt{ffmpeg -i \$file -ss 00:00:03 -t 00:00:15 -vf scale=320:240 \$name}} to the resolution of $320 \times 240$;
    \item  \textbf{cut-kd:} each video was manually cut through \Kdenlive{} (v17.12.3) by keeping $5$ to $7$ seconds and then the video was saved with the \textit{MP4 - the dominating format(H264/AAC)} setting;
    \item \textbf{cut-av:} each video was manually cut through \Avidemux{} (v2.7.4) by keeping $5$ to $7$ seconds and then the video was saved as \textit{copy} and \textit{MP4 Muxer} settings;
    \item \textbf{cut-ap:} each video was manually cut through \APremiere{} Pro CC 2019 by keeping $5$ to $7$ seconds and by saving as H.264 with medium bitrate setting;
    \item \textbf{date change:} each video was manually processed through \Exiftool{} (v11.37) to change the date information within the metadata\footnote{The operation is performed with the command \texttt{exiftool "-AllDates=1986:11:05 12:00:00" \$videos}.}.
\end{itemize}

We considered \Ffmpeg{}, \Exiftool{}, \Avidemux{} and \Kdenlive{} for two main reasons:
\begin{enumerate}
    \item some of them can forge videos in automated way, thus allowing us to create a dataset of tampered videos large enough to obtain statistically significant results;
    \item they allow even a novice to create persuasive forged videos, for instance by cutting specific frames, slowing down or speeding up the streams.
\end{enumerate}
Indeed, some real-world forged videos involved such operations.
The White House suspended access to CNN's Jim Acosta, after he refused to give up the microphone while asking a question about the Russia investigation at a news conference with President Trump.
However, the video reporting the event was possibly speeded up\footnote{\url{https://bit.ly/2vniNi5}}.
Another example is a viral clip of Nancy Pelosi that has been edited to give the impression that the Democratic House speaker was drunk or unwell\footnote{\url{https://bit.ly/2Vx2BGj}}.
We also considered videos manually forged with \APremiere{} a proficient video editing tool that can be used by an expert to produce fake contents.

Furthermore, all the produced contents ($140$ pristine videos and $1260$ tampered ones) were exchanged through different social media platforms:
\begin{itemize}
\item[] \textbf{YouTube videos:} manual upload on YouTube and automated download through \textit{youtube-dl}\footnote{through the command line \texttt{youtube-dl -f mp4 -o "\%(title)s.\%(ext)s" "videos\_list\_link"}};
\item[] \textbf{Facebook videos:} manual upload and download from Facebook with the 'SD' setting.
\item[] \textbf{Tiktok videos:} manual upload and download from TikTok 10.0.0 via a HUAWEI Mate 30 Pro 5G device with the system of EMUI 10.0.0, Android 10.
    Several accounts were used to overcome the uploading limitation.
\item[] \textbf{Weibo videos:} manual upload to Weibo and automated download using Flvcd.0.4.8.1 (http://www.flvcd.com).
\end{itemize}
}
The new dataset thus consists of $7000$ videos (from now on \sk{} Dataset\footnote{\sk{} is available for download from our research group site \url{https://lesc.dinfo.unifi.it/en/datasets}.}).

The container structure, described in \cref{sec:file_format}, is extracted from each video by means of the MP4 Parser library~\cite{mp4parser}.
Note that, due to how the dataset was built, some \textit{value-symbols} are always present in some classes even if they are not relevant for their identification.
For instance, all the cut videos have the same duration even if this is not, per se, relevant for identifying the editing.
As this could lead to artificially higher performance, we manually removed the \textit{value-symbols} associated to the following fields: \texttt{@author}, \texttt{@count}, \texttt{@creationTime}, \texttt{@depth}, \texttt{@duration}, \texttt{@entryCount}, \texttt{@entryCount}, \texttt{@flags}, \texttt{@gpscoords}, \texttt{@matrix}, \texttt{@modelName}, \texttt{@modificationTime}, \texttt{@name}, \texttt{@sampleCount}, \texttt{@segmentDuration}, \texttt{@size}, \texttt{@stuff}, \texttt{@timescale}, \texttt{@version}, \texttt{@width}, \texttt{@height}, \texttt{@language}.

It should also be noted that VISION is composed by several iOS/Android devices and a single Windows phone.
We removed this latter device (\texttt{D17}) from our tests since it is not a representative sample for Windows Phone devices.
For this reason, our approach aims to distinguish between iOS and Android videos only.
This is a negligible limitation given that Windows Phone devices represent less than $0.3\%$ of the mobile devices market~\cite{windows_stats}.

In order to estimate the real-world performance of the proposed method we adopted an exhaustive leave-one-out cross-validation strategy.
We partitioned our dataset in $34$ subsets, each one of them containing pristine, manipulated, and social-exchanged videos belonging to a specific device.
We performed each of the experiments hereby described $34$ times, each time keeping one of the subsets out as test set, and using the remaining $33$ for training our model.
In this way, test accuracies collected after each iteration are computed on videos belonging to an unseen device.
We reported the mean accuracies obtained among all the iterations as confusion matrices.
During the training we assigned to each class a weight inversely proportional to the class frequency.
We used the decision trees algorithms included in  \textit{scikit-learn}~\cite{scikit-learn}, a freely available Python toolkit for machine learning.

\begin{table}[]
 \caption{Balanced accuracies obtained in the basic scenario for each device.}
    \label{tab:q1detailacc}
\begin{tabular}{cc||cc||cc}
Device & \begin{tabular}[c]{@{}c@{}}Balanced \\ Accuracy\end{tabular} & Device & \begin{tabular}[c]{@{}c@{}}Balanced \\ Accuracy\end{tabular} & Device & \begin{tabular}[c]{@{}c@{}}Balanced\\ Accuracy\end{tabular} \\
\hline
D01    & 1.00                                                        & D13    & 1.00                                                        & D26    & 1.00                                                       \\
D02    & 1.00                                                         & D14    & 1.00                                                        & D27    & 1.00                                                       \\
D03    & 1.00                                                         & D15    & 1.00                                                        & D28    & 1.00                                                       \\
D04    & 0.99                                                     & D16    & 1.00                                                        & D29    & 1.00                                                       \\
D05    & 1.00                                                        & D18    & 1.00                                                        & D30    & 1.00                                                       \\
D06    & 1.00                                                        & D19    & 1.00                                                        & D31    & 1.00                                                       \\
D07    & 1.00                                                        & D20    & 1.00                                                        & D32    & 1.00                                                       \\
D08    & 1.00                                                        & D21    & 1.00                                                        & D33    & 1.00                                                       \\
D09    & 1.00                                                        & D22    & 1.00                                                        & D34    & 1.00                                                       \\
D10    & 1.00                                                        & D23    & 1.00                                                        & D35    & 1.00                                                       \\
D11    & 1.00                                                        & D24    & 1.00                                                        &        &                                                         \\
D12    & 0.50                                                     & D25    & 1.00                                                        &        &
\end{tabular}
\end{table}

We trained our method to distinguish between the two classes ``Pristine'' (containing $136$ videos) and ``Tampered'' (containing $1224$ videos).
We obtained a global balanced accuracy of $98.5 \%$, failing only for videos produced by \texttt{D12} (see \Cref{tab:q1detailacc}).
The low accuracy obtained on such a device is reasonably due to the fact that it is the sole Sony smartphone in our dataset.

As a consequence of our strict leave-one-device-out strategy, we have no videos belonging to a Sony device in our training set when \texttt{D12} is tested.
Thus, our algorithm cannot learn the features needed to correctly classify those videos.
This limitation does not always apply as different camera models can exhibit very similar containers.
In such a case, a native video can be correctly classified even if the specific originating device is unavailable in the training set.
This is the case of the LG D290 (\texttt{D04}) that reaches an accuracy of $0.99$.

We also compared our method with two recently proposed algorithms for video integrity~\cite{iuliani2018video,gueera2019we}.
In \cref{tab:comparison} we report the mean global accuracy and the average runtime per fold for the proposed approach and for those two methods.

\subsection{Discussion}
\label{subs:discussion}
\LT{} provides several improvements with respect to the state of the art.
In comparison with \citet{gueera2019we} we achieve a higher accuracy.
This can be reasonably attributed to their use of a smaller feature space; indeed, only a subset of the available pieces of information are extracted without considering their position within the video container.
On the contrary, \LT{} features also include the path from the root to the value, thus providing a stronger discriminating power.
Indeed, this approach allows to distinguish between two videos where the same information is stored in different atoms.
When compared with \citet{iuliani2018video}, \LT{} is capable of obtaining better classification performance with a lower computational cost.
In \citet{iuliani2018video} $O(N)$ comparisons are required since all the $N$ reference-set examples must be compared with a tested video; on the contrary, the cost for a decision tree analysis is $O(1)$ since the output is reached in a constant number of steps.

\begin{figure}
    \centering
    \subfigure[Integrity verification classifier.]{\label{fig:dt_sample_simple}\includegraphics[width=0.8\linewidth]{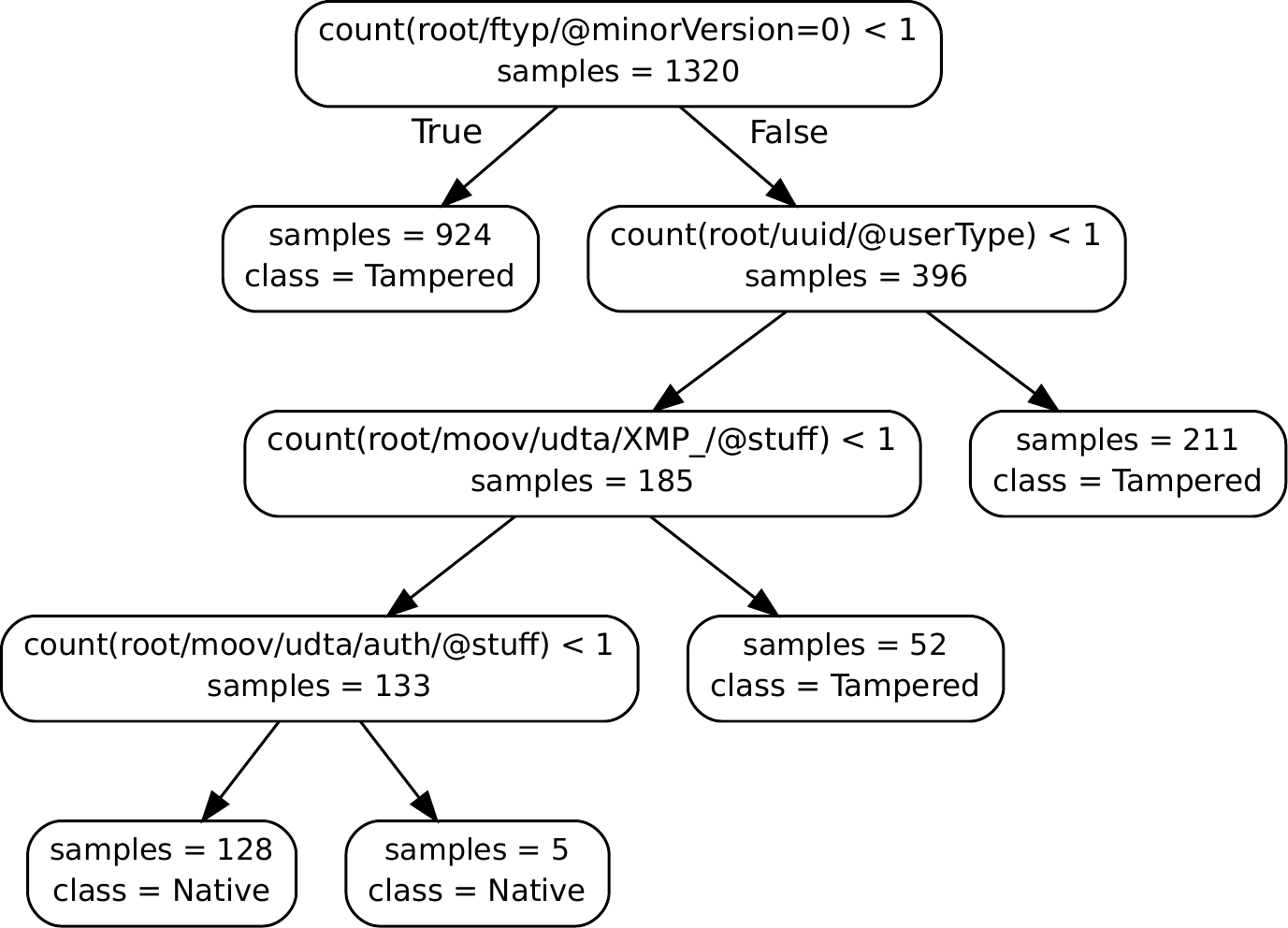}}
    \subfigure[Detail of a blind scenario classifier.]{\label{fig:dt_sample_complex}\includegraphics[width=\linewidth]{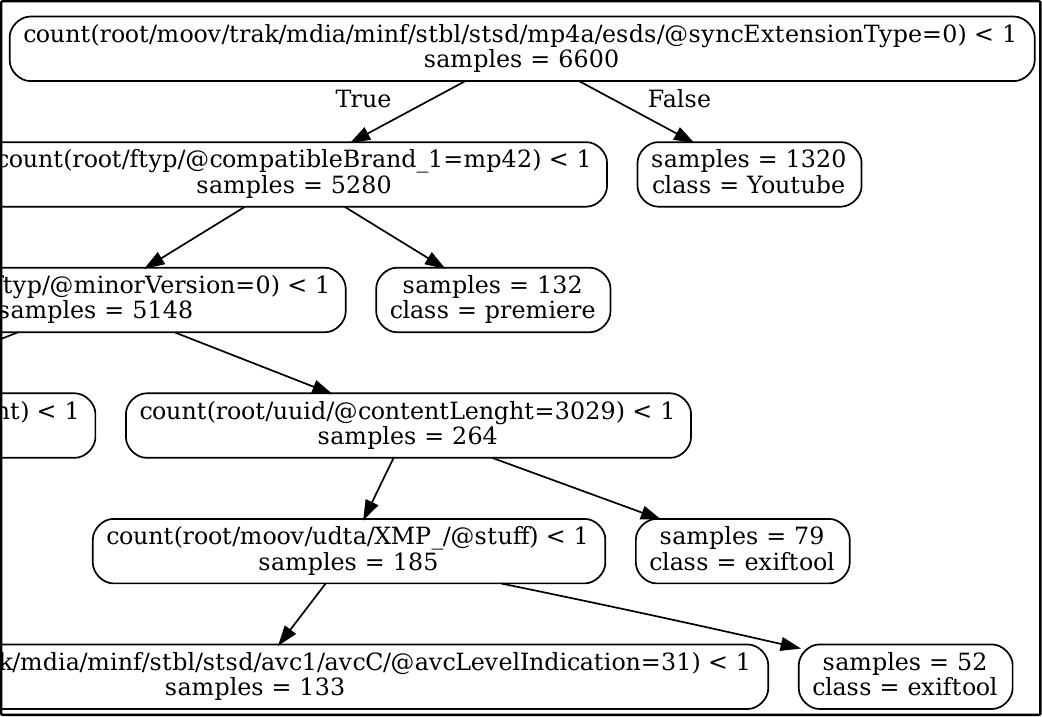}}
    \caption{Pictorial representation of some of the generated decision trees.} \label{fig:dt_sample}
\end{figure}

Furthermore, \LT{} allows a simple explanation for the outcome.
For the sake of example, we report in \cref{fig:dt_sample_simple} a sample tree from the integrity verification experiment: the decision is taken by up to four checks, just based on the presence of the symbols \texttt{ftyp/@minorVersion = 0}, \texttt{uuid/@userType}, \texttt{moov/udta/XMP\_} and \texttt{moov/udta/auth}.
We also report in \cref{fig:dt_sample_complex} a tree from the blind scenario experiment: in this case the tree needs to check the absence of just one atom to classify a YouTube video; at the same time a series of more complex checks are used to assign a video to other classes.
This shows how a single decision tree can handle both easy- and hard-to-classify cases at the same time.
Neither \cite{iuliani2018video} nor \cite{gueera2019we} provide an equivalent feature.
Moreover, \LT{} is equipped with a formal likelihood ratio framework that can estimate the relevance of symbols for specific tasks.
This framework has been used to automatically remove symbols that only contribute to class intra-variability.

\begin{table}[t]
    \centering
    \caption{%
        Comparison of our method with the state of the art.
        Values of accuracy and time are averaged over the 34 folds.
    }
    \label{tab:comparison}
    \begin{tabular}{r||ccc}
                       & \textbf{Balanced accuracy} & \textbf{Training time} & \textbf{Test time} \\ \hline \hline
        Guera et al. \cite{gueera2019we}   & 0.67                      & 347 s         & $<$ 1 s     \\
        Iuliani et al. \cite{iuliani2018video} & 0.85                      & N/A           & 8 s       \\
        \LT{}            & 0.98                      & 31 s          & $<$ 1 s
    \end{tabular}
\end{table}

\section{Manipulation Characterization}
\label{sec:experiments}
We also performed a set of experiments designed to show that the proposed method, as opposed to the state of the art, is also capable of identifying the manipulating software and the operating system of the originating device.
More specifically, we tried to answer the following questions:
\begin{description}
    \item[A] \textbf{Software identification:} Is the proposed method capable of identifying the software used to manipulate a video? If yes, is it possible to identify the operating system of the original video?
    \item[B] \textbf{Integrity Verification on Social Media:} Given a video from a social media platform (YouTube, Facebook, TikTok or WeiBo), can we determine whether the original video was pristine or tampered?
    \item[C] \textbf{Blind scenario:} Given a video that may or may not have been exchanged through a social media platform, is it possible to retrieve some information on the video origin?
\end{description}

\subsection{Software identification}
In this scenario we only analyze videos that either are native, or that have undergone a manipulation.
This time, however, we trained our algorithm to classify which software has been used to tamper the video, if any.
Our classes are thus: ``native'' ($136$ videos), ``\Avidemux{}'' ($136$ videos), ``\Exiftool{}'' ($136$ videos), ``\Ffmpeg{}'' ($680$ videos), ``\Kdenlive{}'' ($136$ videos), and ``\Premiere{}'' ($136$ videos).

\begin{table}[t]
    \centering
    \caption{Confusion matrix for the software identification scenario.}
    \label{tab:q2cm}
    \resizebox{\linewidth}{!}{
        \begin{tabular}{r||cccccc}
            & \textbf{Native} & \textbf{\Avidemux{}} & \textbf{\Exiftool{}} & \textbf{\Ffmpeg{}} & \textbf{\Kdenlive{}} & \textbf{\Premiere{}} \\
            \hline
            \hline
            \textbf{Native}         & 0.97 & -    & 0.03 & -    & -    & -    \\
            \textbf{\Avidemux{}}    & -    & 1.00 & -    & -    & -    & -    \\
            \textbf{\Exiftool{}}    & 0.01 & -    & 0.99 & -    & -    & -    \\
            \textbf{\Ffmpeg{}}      & -    & 0.01 & -    & 0.90 & 0.09 & -    \\
            \textbf{\Kdenlive{}}    & -    & -    & -    & -    & 1.00 & -    \\
            \textbf{\Premiere{}}    & -    & -    & -    & -    & -    & 1.00 \\
        \end{tabular}
    }
\end{table}
In this experiment \LT{} obtained a global balanced accuracy of $97.6 \%$; the detailed results reported in \cref{tab:q2cm} show that the algorithm achieved a slightly lower accuracy in identifying \Ffmpeg{} with respect to the other tools.
This is reasonably due to the fact that \Ffmpeg{} library is used by other software and, internally, by Android devices.

We also trained our algorithm to classify both the editing software used to tamper the video, if any, and the operating system of the device originally used for the acquisition.
The classes for this scenario are: ``Android-native'' ($84$ videos), ``iOS-native'' ($52$ videos), ``Android-avidemux'' ($84$ videos), ``iOS-avidemux'' ($52$ videos), ``Android-exiftool'' ($84$ videos), ``iOS-exiftool'' ($52$ videos), ``Android-ffmpeg'' ($420$ videos), ``iOS-ffmpeg'' ($260$ videos), ``Android-kdenlive'' ($84$ videos), ``iOS-kdenlive'' ($52$ videos), ``Android-premiere'' ($84$ videos), and ``iOS-premiere'' ($52$ videos).

\begin{table*}[t]
    \centering
    \caption{Confusion matrix for the software identification scenario when the OS is taken into account.}
    \label{tab:q2oscm}
    \begin{tabular}{rc||cccccccccccccc}
        & & \multicolumn{2}{c}{\textbf{Native}} & \multicolumn{2}{c}{\textbf{\Avidemux{}}} & \multicolumn{2}{c}{\textbf{\Exiftool{}}} & \multicolumn{2}{c}{\textbf{\Ffmpeg{}}} & \multicolumn{2}{c}{\textbf{\Kdenlive{}}} &\multicolumn{2}{c}{\textbf{\Premiere{}}} \\
        & & Android & iOS & Android & iOS & Android & iOS & Android & iOS & Android & iOS & Android & iOS \\
        \hline
        \hline
        \multirow{2}{*}{\textbf{Native}}        & Android   & 0.95 & -    & -    & -    & 0.05 & -    & -    & -    & -    & -    & -    & -    \\
                                                & iOS       & -    & 1.00 & -    & -    & -    & -    & -    & -    & -    & -    & -    & -    \\
        \multirow{2}{*}{\textbf{\Avidemux{}}}   & Android   & -    & -    & 0.95 & 0.05 & -    & -    & -    & -    & -    & -    & -    & -    \\
                                                & iOS       & -    & -    & -    & 1.00 & -    & -    & -    & -    & -    & -    & -    & -    \\
        \multirow{2}{*}{\textbf{\Exiftool{}}}   & Android   & 0.01 & -    & -    & -    & 0.99 & -    & -    & -    & -    & -    & -    & -    \\
                                                & iOS       & -    & -    & -    & -    & -    & 1.00 & -    & -    & -    & -    & -    & -    \\
        \multirow{2}{*}{\textbf{\Ffmpeg{}}}     & Android   & 0.01 & -    & 0.01 & 0.05 & -    & -    & 0.75 & -    & -    & 0.15 & 0.04 & -    \\
                                                & iOS       & -    & -    & -    & -    & -    & -    & -    & 1.00 & -    & -    & -    & -    \\
        \multirow{2}{*}{\textbf{\Kdenlive{}}}   & Android   & -    & -    & -    & -    & -    & -    & -    & -    & 0.75 & 0.25 & -    & -    \\
                                                & iOS       & -    & -    & -    & -    & -    & -    & -    & -    & 0.38 & 0.62 & -    & -    \\
        \multirow{2}{*}{\textbf{\Premiere{}}}   & Android   & -    & -    & -    & -    & -    & -    & -    & -    & -    & -    & 0.79 & 0.21 \\
                                                & iOS       & -    & -    & -    & -    & -    & -    & -    & -    & -    & -    & 0.37 & 0.63 \\
    \end{tabular}
\end{table*}

A summary of the results obtained by this experiment is reported in \cref{tab:q2oscm}.
Our approach maintains good performance in correctly identifying the editing software.
We notice, however, that the operating system used for videos manipulated with \Kdenlive{} or with \APremiere{} is often misclassified.
At the same time, both those programs are always identified correctly.
This indicates that the container's structure of videos saved by \Kdenlive{} and \APremiere{} is probably reconstructed in a software-specific way.

\subsection{Integrity Verification on Social Media}
In this scenario we tested YouTube, Facebook, TikTok and Weibo videos to determine whether they were pristine or manipulated prior the upload.

\begin{table}[t]
    \centering
    \caption{%
        Performance achieved for integrity verification on social media contents.
        We report for each social network the obtained accuracy, true positive rate (TPR), and true negative rate (TNR).
        All these performance measures are balanced.
    }
    \label{tab:socialdec}
    \begin{tabular}{r||c|cc}
                 & Accuracy & TNR  & TPR  \\ \hline \hline
        Facebook & 0.76     & 0.40 & 0.86 \\
        TikTok   & 0.80     & 0.51 & 0.75 \\
        Weibo    & 0.79     & 0.45 & 0.82 \\
        YouTube  & 0.60     & 0.36 & 0.74 \\
    \end{tabular}
\end{table}

A summary of the results obtained by our method is reported in \cref{tab:socialdec}.
We achieved global balanced accuracies of 0.76, 0.80, 0.79, and 0.60 on Facebook, TikTok, Weibo, and Youtube, respectively.
Such results are characterised by low true negative rates, and thus it cannot be considered effective in this scenario, as many tampered videos are incorrectly classified as pristine.

The poor performance are mainly due to the social media transcoding process that flattens the containers almost independently on the video origin.
As an example, after YouTube transcoding, videos produced by \Avidemux{} and by \Exiftool{} have exactly the same container representation.
We do not know how the videos are processed by the considered platforms due to the lack of public documentation but we can assume that uploaded videos undergo custom/multiple processing.
Indeed, social media videos need to be viewable on a great range of platforms, and thus need to be transcoded to multiple video codecs and adapted for multiple resolutions and bitrates.
Thus, it seems plausible that those operations could discard most of the original container structure.

\subsection{Blind scenario}
In this scenario we considered videos that may or may not have been exchanged through a social media platform and we would like to extract the most complete information possible.
We used all the videos in our dataset and we trained our classifier to distinguish (i) whether the video was downloaded from a social media platform; (ii) whether the video was tampered and, if so, which software was used; (iii) whether the original video belonged to an Android or iOS device.

\begin{table*}[t]
    \centering
    \caption{Confusion matrix for the blind scenario.}
    \label{tab:q4cm}
    \resizebox{\textwidth}{!}{
        \begin{tabular}{rc||cccccccccccccccc}
            & & \multicolumn{2}{c}{\textbf{Native}} & \multicolumn{2}{c}{\textbf{\Avidemux{}}} & \multicolumn{2}{c}{\textbf{\Exiftool{}}} & \multicolumn{2}{c}{\textbf{\Ffmpeg{}}} & \multicolumn{2}{c}{\textbf{\Kdenlive{}}} & \multicolumn{2}{c}{\textbf{\Premiere{}}} & \multirow{2}{*}{\textbf{Facebook}} & \multirow{2}{*}{\textbf{TikTok}} & \multirow{2}{*}{\textbf{Weibo}} & \multirow{2}{*}{\textbf{YouTube}} \\
            & & Android & iOS & Android & iOS & Android & iOS & Android & iOS & Android & iOS & Android & iOS & & & & \\
            \hline
            \hline
            \multirow{2}{*}{\textbf{Native}}        & Android   & 1.00 & -    & -    & -    & -    & -    & -    & -    & -    & -    & -    & -    & -    & -    & -    & -    \\
                                                    & iOS       & -    & 1.00 & -    & -    & -    & -    & -    & -    & -    & -    & -    & -    & -    & -    & -    & -    \\
            \multirow{2}{*}{\textbf{\Avidemux{}}}   & Android   & -    & -    & 0.95 & -    & -    & -    & -    & -    & -    & -    & -    & -    & -    & -    & 0.05 & -    \\
                                                    & iOS       & -    & -    & -    & 1.00 & -    & -    & -    & -    & -    & -    & -    & -    & -    & -    & -    & -    \\
            \multirow{2}{*}{\textbf{\Exiftool{}}}   & Android   & 0.01 & -    & -    & -    & 0.99 & -    & -    & -    & -    & -    & -    & -    & -    & -    & -    & -    \\
                                                    & iOS       & -    & -    & -    & -    & -    & 1.00 & -    & -    & -    & -    & -    & -    & -    & -    & -    & -    \\
            \multirow{2}{*}{\textbf{\Ffmpeg{}}}     & Android   & 0.05 & -    & 0.01 & -    & -    & -    & 0.75 & -    & -    & 0.15 & -    & -    & -    & -    & 0.05 & -    \\
                                                    & iOS       & -    & -    & -    & -    & -    & -    & -    & 1.00 & -    & -    & -    & -    & -    & -    & -    & -    \\
            \multirow{2}{*}{\textbf{\Kdenlive{}}}   & Android   & -    & -    & -    & -    & -    & -    & -    & -    & 0.75 & 0.25 & -    & -    & -    & -    & -    & -    \\
                                                    & iOS       & -    & -    & -    & -    & -    & -    & -    & -    & 0.38 & 0.62 & -    & -    & -    & -    & -    & -    \\
            \multirow{2}{*}{\textbf{\Premiere{}}}   & Android   & -    & -    & -    & -    & -    & -    & -    & -    & -    & -    & 0.80 & 0.20 & -    & -    & -    & -    \\
                                                    & iOS       & -    & -    & -    & -    & -    & -    & -    & -    & -    & -    & 0.37 & 0.63 & -    & -    & -    & -    \\
            \multicolumn{2}{c||}{\textbf{Facebook}}             & -    & -    & -    & -    & -    & -    & -    & -    & -    & -    & -    & -    & 1.00 & -    & -    & -    \\
            \multicolumn{2}{c||}{\textbf{TikTok}}               & -    & -    & -    & -    & -    & -    & -    & -    & -    & -    & -    & -    & -    & 1.00 & - & -    \\
            \multicolumn{2}{c||}{\textbf{Weibo}}                & -    & -    & -    & -    & -    & -    & -    & -    & -    & -    & -    & -    & -    & -    & 1.00 & -    \\
            \multicolumn{2}{c||}{\textbf{YouTube}}              & -    & -    & -    & -    & -    & -    & -    & -    & -    & -    & -    & -    & -    & -    & -    & 1.00 \\
        \end{tabular}
    }
\end{table*}

A summary of the results obtained by our method is reported in \cref{tab:q4cm}.
Even without any prior knowledge of the video origin, we are still able to distinguish between native and tampered videos.
Our method is also able of correctly identifying videos belonging to YouTube, Facebook, TikTok and Weibo, even though in those cases it is not possible to make further claims on the video authenticity.
In most cases we are also able to correctly classify the operating system of the source device.

\section{Conclusions}
\label{sec:conclusions}

In this paper we proposed an efficient forensic method for checking video integrity.
If a manipulation is detected, the proposed method allows to identify the editing software and, in most cases, whether the original video belonged to an Android or iOS device.

This is achieved by exploiting a decision tree classifier applied to a vector based representation of the video container structure, enriched with the likelihood ratio framework that is employed to automatically remove container elements that only contribute to source intra-variability.
The proposed method, in case of tampered videos, is able to characterise the software that performed the manipulation with an accuracy of $97.6 \%$, even when the video is cut without re-encoding.
Except for manipulations performed with \APremiere{} and \Kdenlive{}, the proposed method correctly determines the operating system of the video source device.

As opposed to the state of the art, the proposed method is extremely efficient and can provide a simple explanation for its decisions.
A new experimental dataset of 7000 videos was also created and shared with the research community, including contents generated with five editing tools (\Ffmpeg{}, \Exiftool{}, \APremiere{}, \Avidemux{}, and \Kdenlive{}) and four social media platforms (Facebook, TikTok, Weibo and Youtube).
The current limitation of the method is that a container-based approach can identify whether the video belongs to a social medial platform like YouTube, Facebook, TikTok or Weibo, but it cannot be effectively applied on such contents for authenticity assessment, since the transcoding operation wipes out most of the forensic traces from the video container.
Future works will aim to improve our tool by adding handcrafted features to improve the performance on social media contents.

\section*{Acknowledgment}
The U.S. Government is authorized to reproduce and distribute reprints for Governmental purposes notwithstanding any copyright notation thereon.
The views and conclusions contained herein are those of the authors and should not be interpreted as having to do with the official policies or endorsements, either expressed or implied, of the Air Force Research Laboratory and the Defense Government.

Pengpeng Yang would like to acknowledge the China Scholarship Council, State Scholarship Fund, that supports his joint Ph.D program.

\bibliographystyle{IEEEtranN}
\bibliography{paper}

\end{document}